\documentclass[acs,jacs,reprint,showpacs,superscriptaddress]{revtex4-1}
\usepackage{amsfonts,mathrsfs,amsmath,amsthm,amssymb,bbold}
\usepackage{epsfig}
\usepackage{bm,mathrsfs}
\usepackage{lineno,hyperref}
\usepackage{amsmath}
\usepackage{color}
\usepackage{xcolor,colortbl} 
\usepackage{soul}
\usepackage{afterpage}
\usepackage{capt-of}
\usepackage{makecell}
\usepackage{graphicx}
\setcellgapes{4pt}

\newcolumntype{P}[1]{>{\centering\arraybackslash}p{#1}}

\usepackage{amsmath} % or simply amstext

\begin{document}
\sloppy
\clearpage

\title[mode  = title]{Anisotropic magnetocaloric effect of CrI$_{3}$: A theoretical study}% Force line breaks with \\

%
\author{Hung Ba Tran}
%\thanks{Electronic address: tran.h.ag@m.titech.ac.jp}
\address{Laboratory for Materials and Structures, Institute of Innovative Research, Tokyo Institute of Technology, Midori-ku, Yokohama 226-8503, Japan}
\address{Quemix Inc., 2-11-2 Nihombashi, Chuo-ku, Tokyo 103-0027, Japan}
\address{Institute of Scientific and Industrial Research, Osaka University, 8-1 Mihogaoka, Ibaraki, Osaka 567-0047, Japan}
%\address{Department of Precision Science and Technology, Graduate School of Engineering, Osaka University, 2-1 Yamada-oka, Suita, Osaka 565-0871, Japan}
%\affiliation{Laboratory for Materials and Structures, Institute of Innovative Research, Tokyo Institute of Technology, Midori-ku, Yokohama 226-8503, Japan}

%
\author{Hiroyoshi Momida}
\address{Institute of Scientific and Industrial Research, Osaka University, 8-1 Mihogaoka, Ibaraki, Osaka 567-0047, Japan}

%
%\author{Kazunori Sato}
%\address{Divisions of Materials and Manufacturing Science, Graduate School of Engineering, Osaka University, 2-1 Yamada-oka, Suita, Osaka 565-0871, Japan}
%\address{Center for Spintronics Research Network, Osaka University, Toyonaka, Osaka 560-8531, Japan}

%
\author{Yu-ichiro Matsushita}
\address{Laboratory for Materials and Structures, Institute of Innovative Research, Tokyo Institute of Technology, Midori-ku, Yokohama 226-8503, Japan}
\address{Quemix Inc., 2-11-2 Nihombashi, Chuo-ku, Tokyo 103-0027, Japan}

%
%\author{Yukihiro Makino}
%\address{Daikin Industries, LTD, 1-1 Nishi-Hitotsuya, Settsu, Osaka 566-8585, Japan}

%
\author{Koun Shirai}
\address{Institute of Scientific and Industrial Research, Osaka University, 8-1 Mihogaoka, Ibaraki, Osaka 567-0047, Japan}

\author{Tamio Oguchi}
%\thanks{Electronic address: oguchi@sanken.osaka-u.ac.jp}
\address{Institute of Scientific and Industrial Research, Osaka University, 8-1 Mihogaoka, Ibaraki, Osaka 567-0047, Japan}
\address{Center for Spintronics Research Network, Osaka University, Toyonaka, Osaka 560-8531, Japan}

\date{\today}% It is always \today, today,
    % but any date may be explicitly specified

\begin{abstract}
CrI$_{3}$ is considered to be a promising candidate for spintronic devices and data storage. We derived the Heisenberg Hamiltonian for CrI$_{3}$ from density functional calculations using the Liechtenstein formula. Moreover, the Monte--Carlo simulations with the Sucksmith--Thompson method were performed to analyze the effect of magnetic anisotropy energy on the thermodynamic properties. Our method successfully reproduced the negative sign of isothermal magnetic entropy changes when a magnetic field was applied along the hard plane. We found that the temperature dependence of the magnetocrystalline anisotropy energy is not negligible at temperatures slightly above the Curie temperature. We clarified that the origin of this phenomenon is attributed to anisotropic magnetic susceptibility and magnetization anisotropy. The difference between the entropy change of the easy axis and the hard plane is proportional to the temperature dependence of the magnetic anisotropy energy, implying that the anisotropic entropy term is the main source of the temperature dependence of the free energy difference when magnetizing in a specific direction other than the easy axis. We also investigated the magnetic susceptibility that can be used for the characterization of the negative sign of the entropy change in the case of a hard plane. The competition of magnetocrystalline anisotropy energy and external magnetic field at low temperature and low magnetic field region causes a high magnetic susceptibility as the fluctuation of magnetization. Meanwhile, the anisotropy energy is suppressed at a sufficient magnetic field applied along the hard axis, the magnetization is fully rotated to the direction of the external magnetic field.
\end{abstract}

\maketitle

%\tableofcontents

\section{\label{sec:level1}Introduction}
%para1: 
The development of Si-CMOS technology is a backbone of electronic devices such as computers and smartphones\cite{FrankIEEE2001,DennardIEEE1974}. The decrease in the transistor size has many advantages, such as high efficiency and low power consumption. However, the size reduction is a difficult task when it is close to the critical size\cite{ThomasNE2018}. Researchers are attempting to identify alternative methods to traditional computers, such as quantum and neuromorphic computers, which have many advantages over traditional computing hardware \cite{LeuenbergerPE2001,SenguptaAPR2017,GrollierIEEE2016,AwschalomScience2013}. Spintronic devices are promising as an essential part of future computing devices because they utilize the spin and charge of electrons\cite{HirohataJMMM2020}. Recent studies on spin logic gates using magnetic materials with layered structures have provided an opportunity for future computing hardware with low power consumption and high speed\cite{Ahn2DMA2020}. 

%para2:
The recent discovery of 2D materials also has potential applications in spintronic devices and data storage\cite{Ahn2DMA2020,GibertiniNN2019}. They show a possible way to realize a device with low power consumption, high efficiency, compactness, and multi-functionality\cite{Ahn2DMA2020,GibertiniNN2019}. Among several 2D materials, van-der-Waals magnetic materials with layered structures have attracted considerable attention because they are easy to manufacture and have some unique properties\cite{Ahn2DMA2020,GibertiniNN2019,HuangNN2018}. Layered magnetic materials have potential for application as data storage and memory devices, where the layer's size can be reduced to a few nanometers\cite{Ahn2DMA2020,GibertiniNN2019,HuangNN2018}. In this case, magnetocrystalline anisotropy energy (MAE) and interlayer exchange coupling are vital for controlling the magnetic properties at finite temperatures. The MAE is the energy required to change the direction of crystal magnetization from the preferred axis (easy axis) to another hard axis\cite{CallenJPCS1966, AsselinPRB2010}. Understanding the behavior and origin of the temperature dependence of MAE provides many opportunities for controlling magnetic devices \cite{CallenJPCS1966,AsselinPRB2010}. Recent studies on perpendicular magnetic anisotropy in 2D materials have shown a clue to fabricate low-power consumption devices with fast responses\cite{Ahn2DMA2020,GibertiniNN2019}. Accurate prediction of MAE and its temperature dependence will enable the design of data storage devices in spintronics\cite{CallenJPCS1966, AsselinPRB2010}.

%para3:
CrI$_{3}$ is a compound of chromium and iodine, which is a magnetic insulator of 2D materials \cite{Ahn2DMA2020,GibertiniNN2019,HuangNN2018}. Bulk CrI$_{3}$ has two crystal structures: the low-temperature phase rhombohedral with space group $R\bar{\rm 3}$ and the high-temperature phase monoclinic with space group $C2/m$\cite{HandyJACS1952,MorosinJCP1964,PolliniSSC1998,WangJP2011}. The material is ferromagnetic (FM) below the Curie temperature of 61 K and has uniaxial magnetic anisotropy with the easy axis being the $c$ axis and the hard plane being the $ab$ plane\cite{HandyJACS1952,MorosinJCP1964,PolliniSSC1998,WangJP2011,McGuireCM2015,LiuPRB2018,McGuireCM2015,LiuPRB2018}. CrI$_{3}$ has been experimentally and theoretically studied since it was applied to successfully manufacture monolayer and multilayer structures from bulk\cite{HandyJACS1952,MorosinJCP1964,PolliniSSC1998,WangJP2011,McGuireCM2015,LiuPRB2018,HuangNN2018}. Although the MAE of bulk CrI$_{3}$ was calculated using density functional theory, the temperature dependence of MAE is not well understood yet\cite{McGuireCM2015,LiuPRB2018,GudelliNJP2019}. The anisotropic magnetocaloric effect, which is the response of a magnetic anisotropy material with an external magnetic field, has been reported\cite{LiuPRB2018}. The peak-like magnetization curves of CrI$_{3}$ observed when applying a medium external magnetic field along the hard plane are quite unusual because the magnetization increases when the temperature increases\cite{McGuireCM2015,LiuPRB2018}. In addition, alternation of the sign from negative to positive for isothermal magnetic entropy was observed when the magnetic field was applied along the hard plane at low temperatures, which remains an open question\cite{LiuPRB2018}. 

%para4: In this work
In this study, the electronic structure and magnetic and magnetocaloric properties of bulk CrI$_{3}$ with a rhombohedral structure were investigated using first-principles calculations and Monte--Carlo simulations. The electronic structure and magnetic exchange coupling constants of several magnetic configurations were calculated. The dependence of Curie temperature on the number of nearest neighbors was estimated using the mean-field approximation (MFA) and Monte--Carlo method. In addition, the magnetization versus magnetic field and temperature ($M$--$H$ and $M$--$T$) curves, the temperature dependence of MAE and the anisotropic isothermal magnetic entropy change were compared with the experimental data from previous studies\cite{McGuireCM2015,LiuPRB2018}. To understand the origin and characteristics of negative signs in the entropy change, we investigated the magnetic susceptibility and the relation between MAE and anisotropic magnetic entropy change. Our studies give a comprehensive understanding of the anisotropic magnetocaloric effect with magnetic anisotropy, where the temperature and magnetic field dependence of MAE is a key factor on the thermodynamic properties. 

\section{\label{sec:level2}Methodology}
The electronic structure of bulk CrI$_{3}$ was calculated using DFT with the all-electron full-potential linearized augmented plane wave method as implemented in the HiLAPW package\cite{hilapw}. The experimental lattice structure was assumed for the rhombohedral ($R\bar{\rm 3}$) structure\cite{McGuireCM2015,LiuPRB2018}. The atomic positions were optimized by minimizing the residual atomic forces to be less than 1 mRy/Bohr. The generalized gradient approximation (GGA--PBE) was used for exchange and correlation\cite{GGA-PBE}. The cutoff energy was set as 20 Ry for the wavefunctions and 160 Ry for the potential. The magnetic anisotropy energy was evaluated as the energy difference by rotating the quantization axis with the spin-orbit coupling (SOC) in the second variation step. The convergence of MAE is achieved using a number of \textit{k}-point meshes of 18$\times$18$\times$18 for a primitive cell with rhombohedral axis. 

The magnetic exchange coupling constants between atoms in bulk CrI$_{3}$ were calculated by DFT using the Liechtenstein formula within the Korringa-Kohn-Rostoker method as implemented in the Machikaneyama package\cite{LiechtensteinJMMM1987,AkaiKKR,AkaiPRB1993}:

\begin{equation}
J_{ij}=\frac{1}{4\pi}\mathrm{Im}\int^{E_F}dE\mathrm{Tr}\left \{ \Delta t_i(E)T_{ij}^\uparrow\Delta t_j(E)T_{ij}^\downarrow \right \},
\label{Eq1}
\end{equation}

\noindent where $T_{ij}^\uparrow$ and $T_{ij}^\downarrow$ are the off-diagonal scattering path operators for spin-up and spin-down between atomic sites $i$ and $j$, respectively, and $\Delta t_i(E)=t_{i}^{\uparrow}-t_{i}^{\downarrow}$ with the single site $\textit{t}$-matrix $t_{i}^{\sigma}$ for spin $\sigma$ at site \textit{i}.

The classical Heisenberg model with uniaxial anisotropy is used as:

\begin{equation}
H_{\rm Heis}=-\sum_{<ij>}J_{ij}^{m}\overrightarrow{S_{i}}\overrightarrow{S_{j}}-\sum_{i}k_{\rm u}(\overrightarrow{e_{u}}\overrightarrow{S_{i}})^{2}-g\mu _{\rm B}\sum_{i}\overrightarrow{H_{\rm ext}}\overrightarrow{S_{i}},
\label{Eq2}
\end{equation}  
 
\noindent where $g$ is the gyromagnetic ratio and $\mu_{\rm B}$ is the Bohr magneton. $k_{\rm u}$ denotes the uniaxial anisotropy constant. The first term expresses the exchange interactions between spins at sites \textit{i} and \textit{j}. A spin tends to be parallel to the neighboring site when the magnetic exchange coupling constant $J_{ij}^{m}$ is positive. A spin prefers to be antiparallel with the neighboring spin for negative $J_{ij}^{m}$. The second term is the interaction between the spin at site \textit{i} and the uniaxial anisotropy. $e_{\rm u}$ is the direction of the easy axis in the case of a positive $k_{\rm u}$. The last term denotes the interaction of the spin at site $i$ with the external magnetic field. In an external magnetic field, spins tend to be parallel to the direction of the magnetic field to minimize energy. The size of the cell in the Monte--Carlo simulation in CrI$_{3}$ with hexagonal type is 20$\times$20$\times$20 unit cells containing 192000 atoms. The number of Monte--Carlo steps is 2000000, and the first 1000000 steps are discarded. The Metropolis algorithm was used to achieve thermal equilibrium in the Heisenberg model.

The magnetic susceptibility was calculated as\cite{HBTJAC2021,HBTSSC2021}

 \begin{equation}
\chi = \frac{\partial M}{\partial H} = \frac{(\left \langle M^{2} \right \rangle -\left \langle M \right \rangle^{2})}{k_{\rm B} T}
\label{Eq3}
\end{equation}

\noindent where $M$ is magnetization, $H$ is the magnetic field, $k_{\rm B}$ is the Boltzamnn constant, and $T$ is temperature.

%Figure 1
\begin{figure*} 
\centering
\includegraphics[width=16.0cm]{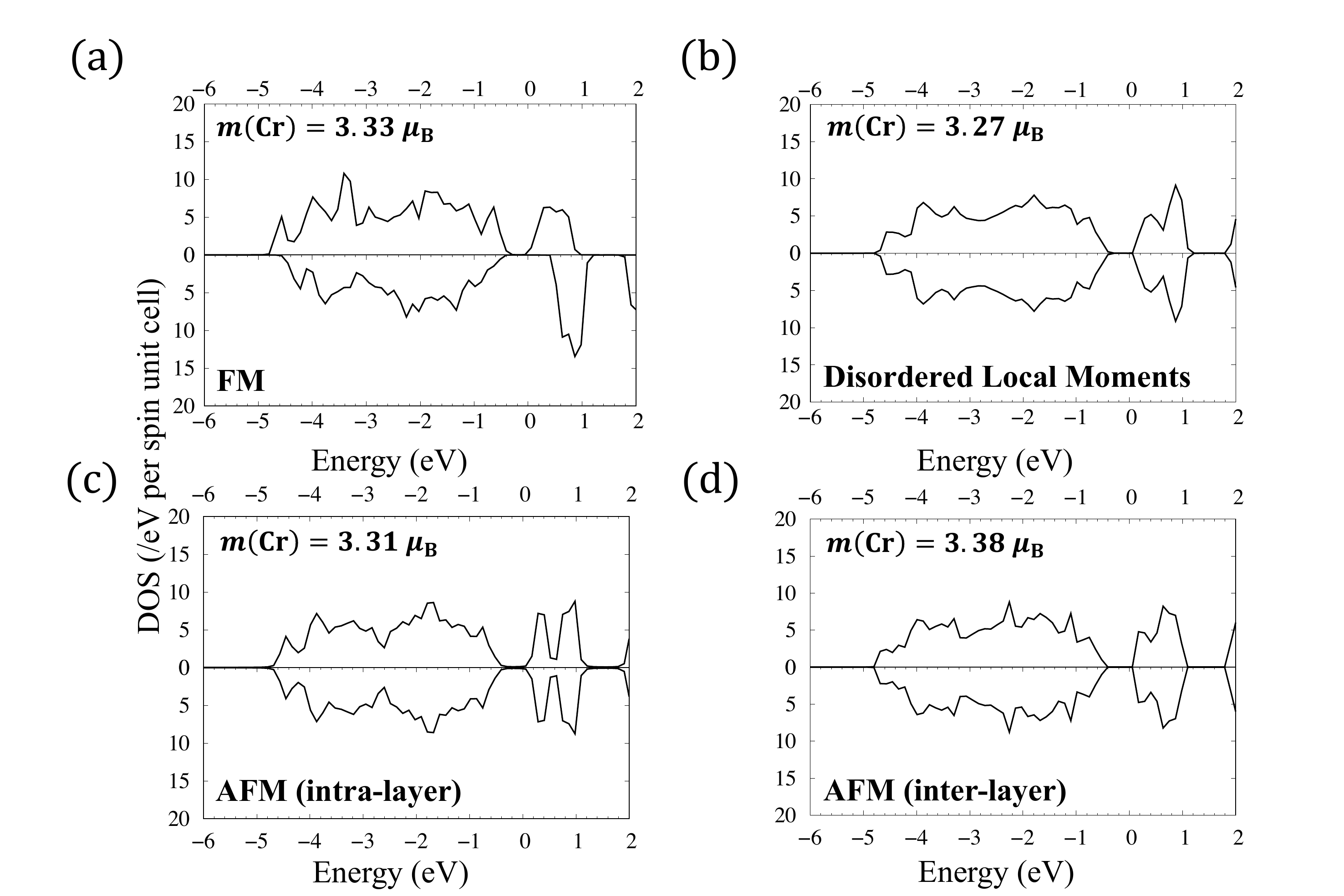} 
\caption{The total density of states of rhombohedral CrI$_{3}$ with various magnetic configurations: (a) ferromagnetic, (b) disordered local moment, (c) antiferromagnetic with up and down magnetic moment in the same layer (AFM-intra-layer), and (d) antiferromagnetic with up and down magnetic moment in a different layer (AFM-inter-layer). } 
\label{FIG1}
\end{figure*}

The MAE was estimated by integrating the magnetization versus the magnetic field ($M$--$H$) curves as

\begin{equation}
{\rm MAE}=\int_{0}^{H_{\rm ext}} \{ M(H\parallel{\rm easy\;axis})-M(H\parallel{\rm hard\;axis})\}dH
\label{Eq4}
\end{equation}

\noindent where integration is considered until the saturation magnetization is achieved like the magnetic field equal to 7 T. The MAE is estimated here as the area between the magnetization curves when applying magnetic field along the easy axis and hard plane in the integration. The temperature dependence of the MAE can be estimated because the strength of the magnetization and anisotropy field at finite temperatures is considered in the value and direction of magnetization curves. The area between the two magnetization curves---one is when the magnetic field along the easy axis and the other is the magnetic field along the hard plane---is considered as the energy cost for magnetization in a certain direction other than the preferred axis.

The isothermal magnetic entropy change is estimated using the Maxwell relation as\cite{HBTJAC2021,HBTSSC2021}

\begin{equation}
\begin{split}
& \Delta S_{\rm M}(H_{\rm ext},T) =\int_{0}^{H_{\rm ext}}\left (\frac{\partial M(H,T) }{\partial T} \right ), \\
& \cong \sum_{j=0}^{N}\frac{M(H_j,T+\Delta T)-M(H_j,T-\Delta T)}{2\Delta T}\Delta H,
\end{split}
\label{Eq5}
\end{equation}

\noindent The isothermal magnetic entropy change is obtained by integrating with the fine mesh in the temperature and external magnetic field. $\Delta T$ and $\Delta H$ are equal to 2 K and 0.2 T, respectively. Owing to the MAE, the isothermal magnetic entropy changes strongly depend on the direction of the applied magnetic field.

\section{\label{sec:level3}Results and discussion}
\subsection{Electronic structure and magnetic exchange coupling constants}

%para 5
The electronic structures of bulk CrI$_{3}$ with ferromagnetic (FM), disordered local moment (DLM) or paramagnetic (PM), and two antiferromagnetic (AFM) configurations are shown in FIG. \ref{FIG1}. All electronic structures are semiconducting with a similar local magnetic moment of the Cr-atom $3.33$ ($\mu_{\rm B}$/atom) as the Cr$^{3+}$ state. The magnetic moment of I is $-0.13$ ($\mu_{\rm B}$/atom) in the FM state, which becomes zero in the other magnetic configurations by symmetry. Our results for the electronic structure of DLM show a semiconducting structure, which is different from previous studies\cite{McGuireCM2015, LiuPRB2018}, which have showed a metallic structure. The discrepancy may be caused by ignoring spin polarization in the previous studies\cite{McGuireCM2015, LiuPRB2018}. The MAE of the FM state was calculated to be 0.46 (meV/Cr-atom) between the easy axis ($c$-axis) and hard plane ($ab$-plane) in the hexagonal coordinates. This agrees with the previous theoretical work in which the MAE was found to be 0.47 (meV/Cr-atom) using the same exchange-correlation function\cite{GudelliNJP2019}. The source of the strong MAE originates from the hybridization between the $d$-states of Cr and the $p$ states of I with a large SOC. The value of MAE at a low temperature is 0.26 (meV/Cr-atom) in an experimental study by integrating the $M$--$H$ curves, which is smaller than the value of first-principles calculations\cite{McGuireCM2015}. The MAE of theoretical studies often tend to be higher than that of experimental studies and the value of the anisotropy field at low temperatures will be higher than that of the experimental studies\cite{McGuireCM2015,LiuPRB2018}.  

%Figure 2
\begin{figure*} 
\centering
\includegraphics[width=16.0cm]{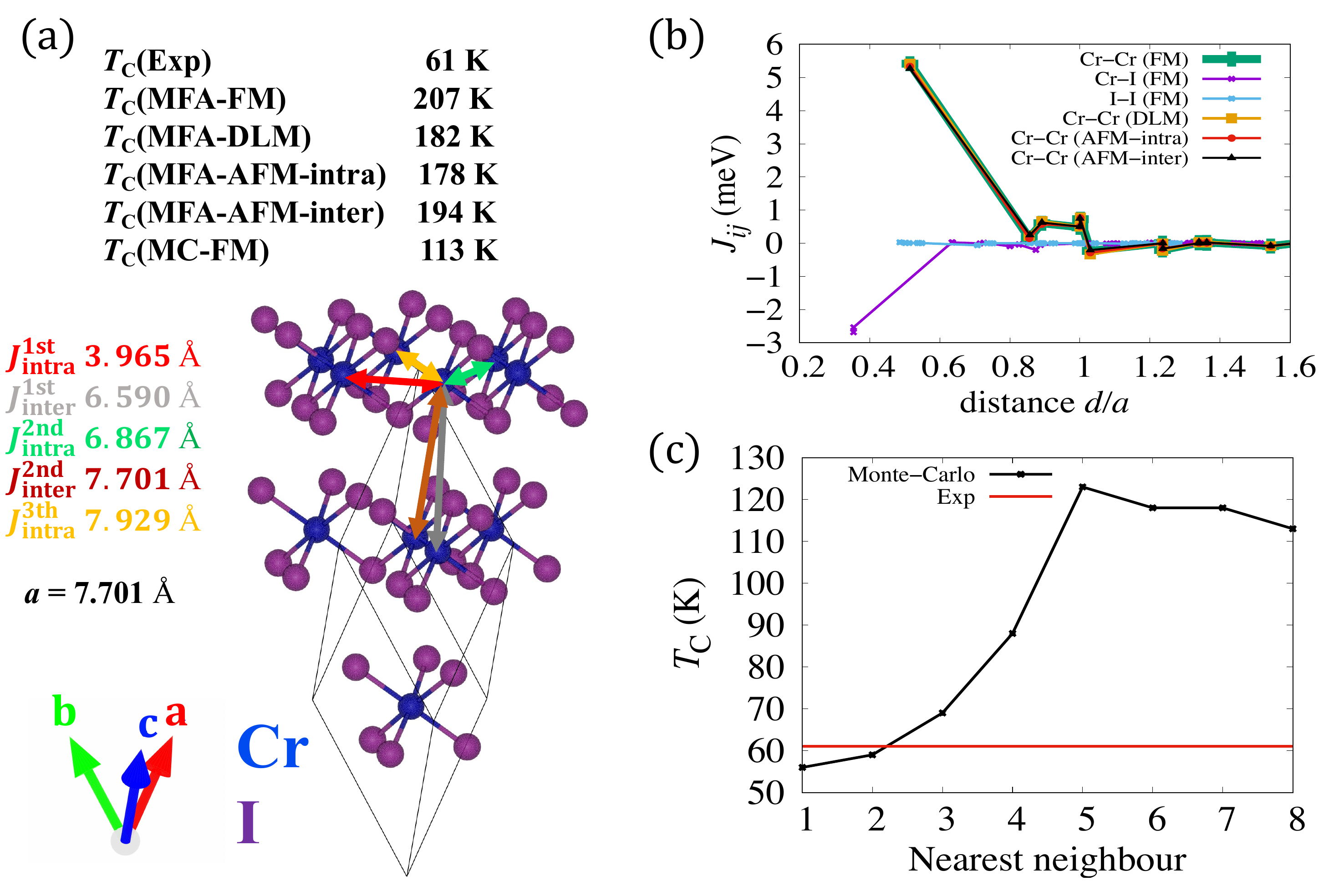} 
\caption{(a) The crystal structure of CrI$_{3}$ with several Cr--Cr pairs and the value of the Curie temperature of experimental  and the present study, which is obtained using MFA and Monte--Carlo simulations. (b) Magnetic exchange coupling constants as a function of distance over the lattice constant of the crystal structure in (a). (c) Curie temperatures of experimental work (red line) and Monte--Carlo simulations depend on the range of the nearest neighbor sites of Cr--Cr pairs (black line). } 
\label{FIG2}
\end{figure*}

%para 6
The magnetic exchange coupling constants of several magnetic configurations and their Curie temperatures in the present calculations are shown in FIG. \ref{FIG2} (a) and (b). Because only the FM phase has a finite local magnetic moment at the I atom, the magnetic exchange coupling constants of the Cr-I and I-I pairs are negligible in the DLM (PM) and AFM states. The large negative coupling constants of the first nearest neighbor of the Cr-I pair are consistent with the values of the magnetic moments of the Cr and I atoms because it minimizes the total energy of the FM state in the Heisenberg model. Moreover, the magnetic exchange coupling constants of the Cr--Cr pairs are not sensitive to the choice of magnetic configurations because the magnetic moment of the Cr-atom is localized. The coupling constant of the first nearest neighbor between Cr atoms in the same layer, which has the largest positive value. This is the main contribution to the intra-layer FM order. In addition, the coupling constant of the second nearest neighbor in the intra-layer is a small positive, while the third one is negative (the distance is slightly larger than one lattice constant). Meanwhile, the coupling constants of the first and second nearest neighbor Cr--Cr pairs in the different layers are small and positive, which is the origin of the stable ferromagnetic order observed in bulk CrI$_{3}$. 

%para 7
Coupling constants up to 15 \AA\;are considered to primarily determine the Curie temperature within MFA. However, MFA usually overestimates the Curie temperature compared with experiment. The Curie temperature obtained by MFA was approximately three times larger than the experimental value\cite{McGuireCM2015, LiuPRB2018}. The Monte--Carlo method was used to obtain a more accurate Curie temperature. In the Monte--Carlo method, we consider up to eight orders of the nearest neighbor of the Cr--Cr pair in the FM state. Although the value is much smaller than the value of MFA, it is still higher than the value of the experimental works. This contradicts previous theoretical studies, which show very good agreement with the experimental value in the Curie temperature\cite{OlsenMRSC2019,LuPRB2019,LiuPCCP2016}. In the previous studies, the magnetic exchange coupling constants were estimated by considering the total energy difference of several magnetic configurations, which makes it difficult to consider the long-range order of the coupling constant\cite{OlsenMRSC2019,LuPRB2019,LiuPCCP2016}. The Curie temperature of the Monte--Carlo method as a function of the range of the magnetic exchange coupling constants is shown in FIG. \ref{FIG2} (c). The estimation of the Curie temperature strongly depends on the range of magnetic interactions in the simulation. Less than the fourth nearest neighbor of Cr-Cr pairs leads to an accidentally good agreement with the experimental value. Although the Curie temperature was overestimated, the long-range interaction of Cr-Cr pairs was considered in the present study. Other effects, such as phonon need to be considered to achieve a more accurate value for the Curie temperature of CrI$_{3}$ in the presence of long-range magnetic interactions \cite{TanakaCM2020}.

\subsection{Finite temperature MAE and anisotropic magnetocaloric effect}

%Figure 3
\begin{figure} 
\centering
\includegraphics[width=8.4cm]{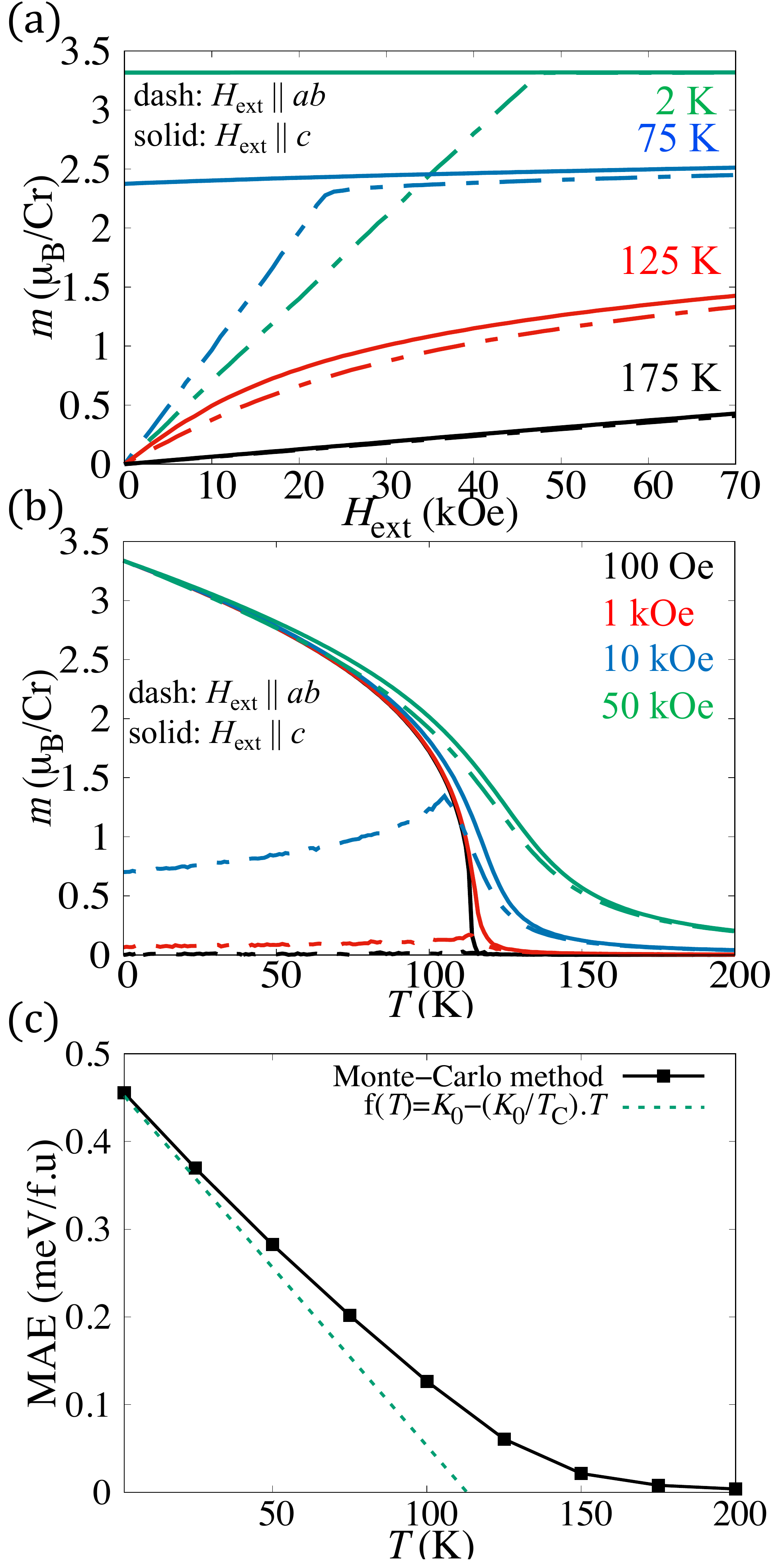} 
\caption{(a) $M$--$H$ curves of several temperatures with the magnetic field along the hard plane (the dashed line) and easy axis (the solid line). (b) Magnetization curves versus temperature for various magnetic field strengths, where the magnetic field is along the hard plane (the dashed line) and easy axis (the solid line). (c) Temperature dependence of MAE by integrating the $M$--$H$ curves in Monte--Carlo simulations.} 
\label{FIG3}
\end{figure}

%para 8
The $M$--$H$ curves at low temperatures, intermediate temperatures, slightly above the Curie temperature, and high temperatures are shown in FIG. \ref{FIG3} (a). The area between the easy axis and hard plane is integrated to estimate MAE, where the effect of temperature on the magnetization and magnetic anisotropy is considered. At a low temperature (2 K) and zero external magnetic field, the magnetization is along the easy axis and is perpendicular to the hard plane. However, the projection of magnetization along the direction of the magnetic field increases when a higher external magnetic field is applied along the hard plane. The magnetization was fully rotated to the hard plane if the magnetic field was sufficient (higher than the anisotropy field). In contrast, the magnetization anisotropy, defined as the difference in the saturation magnetization of the easy axis and the hard plane at the zero external magnetic fields, appears at an intermediate temperature. In addition, the slopes of the two magnetization curves (dashed and solid red lines in FIG. \ref{FIG3} (a)) which denote the differential magnetic susceptibility are similar when the external magnetic field is larger than the anisotropy field.  As a result of magnetization anisotropy and differential magnetic susceptibility at intermediate temperatures, the estimation of MAE strongly depends on the strength of the external magnetic field. If temperature is slightly higher than the Curie temperature, the magnetization at the zero external magnetic field equals zero. However, the application of an external magnetic field can lead to a finite magnetization. The differential magnetic susceptibility of the easy axis is higher than that of the hard plane in the case of a paramagnetic state. This is consistent with the experimental work\cite{McGuireCM2015}. Note that we need to consider the relative temperature because the Curie temperature of our result is higher than the experimental value\cite{McGuireCM2015}. Moreover, the anisotropic magnetic susceptibility decreased rapidly with increasing temperature. At high temperatures, the disappearance of magnetization anisotropy and anisotropic magnetic susceptibility led to a negligible MAE.

%Figure 4
\begin{figure*} 
\centering
\includegraphics[width=16.0cm]{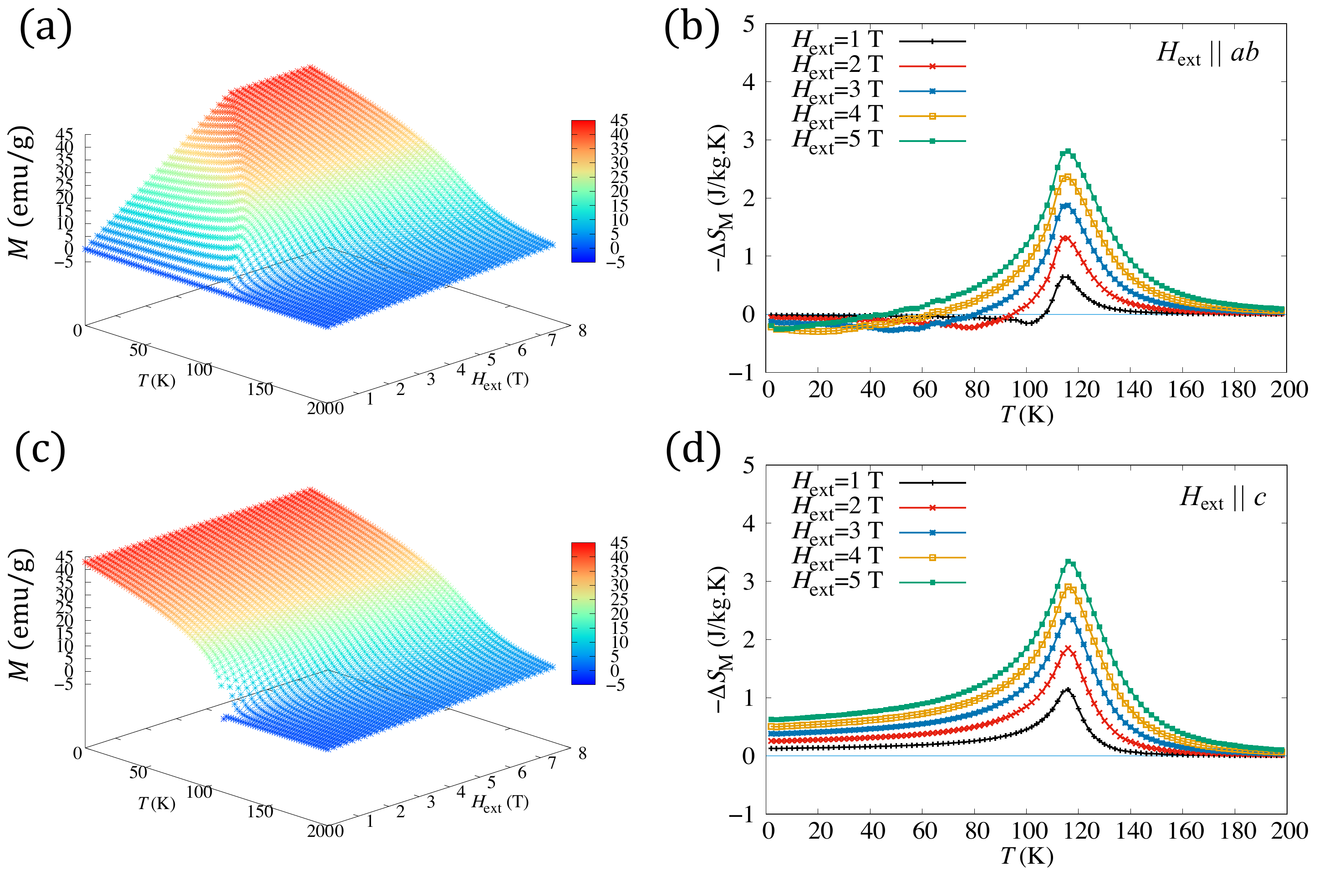} 
\caption{The isothermal magnetization depends on the temperature and strength of the magnetic field, with the direction of field along with the hard plane (a) and easy axis (c). The isothermal magnetic entropy changes as a function of the temperature and magnetic field strength for the hard plane (b) and easy axis (d).} 
\label{FIG4}
\end{figure*}

%Figure 5
\begin{figure} 
\centering
\includegraphics[width=8.0cm]{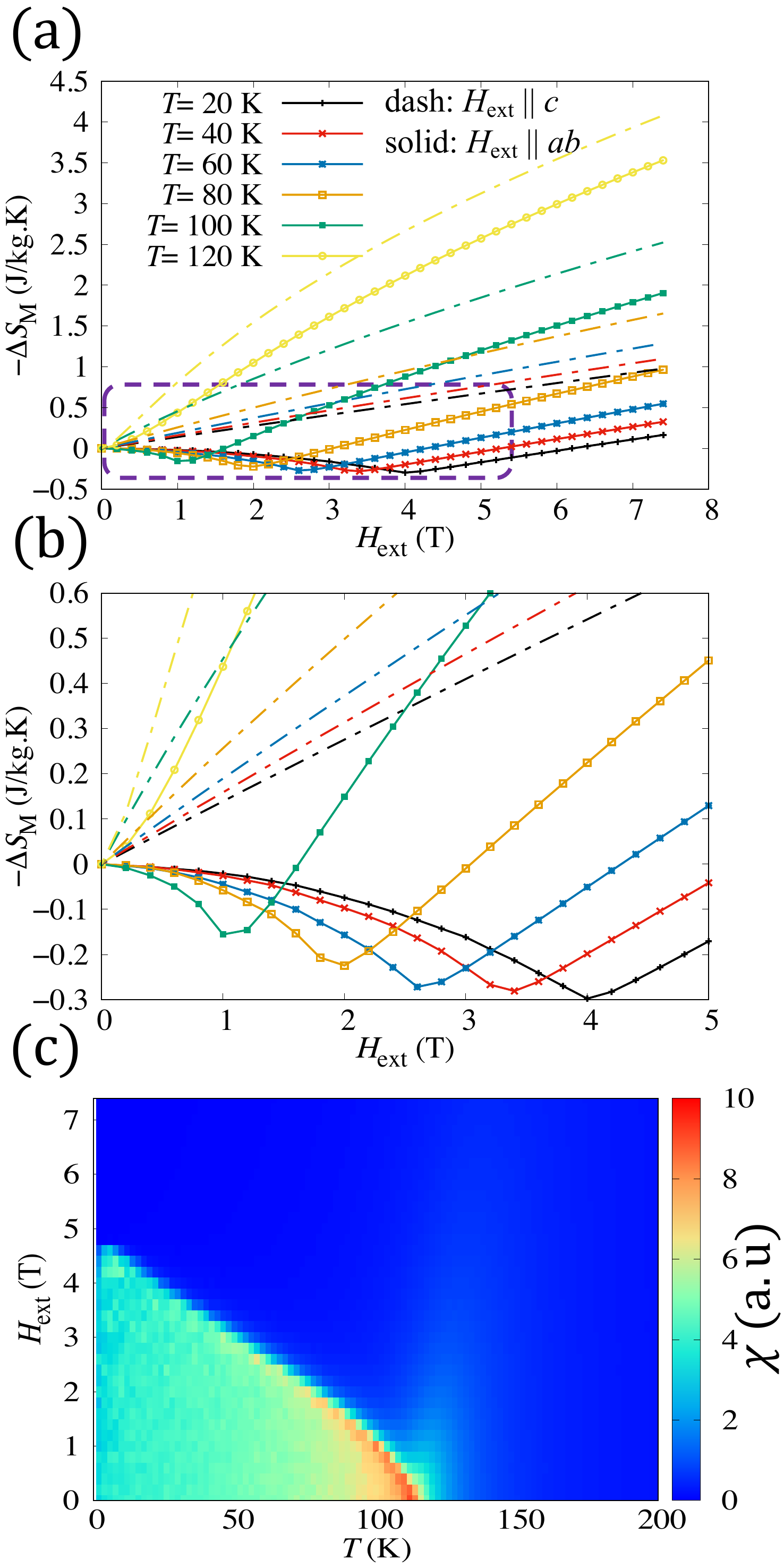} 
\caption{The entropy change depends on the strength of the magnetic field with the magnetic field along the hard plane (solid line) and the easy axis (the dashed line) at various temperatures for large range (a) and small range (b). Note that (b) is the inset region of (a) with a dashed purple line. (c) The magnetic susceptibility depends on the temperature and magnetic field. } 
\label{FIG5}
\end{figure}

%para 9
The temperature dependence of the magnetization curves when a magnetic field is applied along the easy axis and hard plane is shown in FIG. \ref{FIG3} (b). At a weak external magnetic field (100 Oe), the magnetization is mainly along the easy axis when the temperature is lower than the Curie temperature. If the temperature increases, there is a second-order magnetic phase transition (FM-PM) characteristic of the Heisenberg model. Although the magnetization curve of the hard axis agrees well with the experimental results, the magnetization curve of the easy axis in our results is much higher than that of the experimental one\cite{McGuireCM2015,LiuPRB2018}. When the magnetic field is equal to 10 kOe along the easy axis, the magnetization at 0 K in the experimental work is 1.2 ($\mu_{\rm B}$/Cr-atom), while our result is 3.33 ($\mu_{\rm B}$/atom), which indicates saturation magnetization. There are multiple magnetic domains in experimental studies if the magnetic field is insufficient to form a single magnetic domain. However, the size of the simulation box in our results contains only a single magnetic domain. Increasing the external magnetic strength to 1 kOe and 10 kOe leads to a large enhancement in the hard plane.

%para9.1
Peak-like magnetization curves appear for an intermediate magnetic field (the blue dashed line in FIG. \ref{FIG3} (b)). The temperature value of the peak is called $T^{*}$, which is slightly smaller than the Curie temperature and decreases with increasing strength of the magnetic field\cite{McGuireCM2015,LiuPRB2018}. It is only observed in the case of the hard plane and when an intermediate magnetic field is applied. Its origin is the decrease in the anisotropy field when the temperature increases, so the magnetization projection in the direction of the magnetic field increases when the temperature increases. This can be seen as the case of the crossing of the dashed lines of 2 K and 75 K in FIG. \ref{FIG3} (a). The magnetization at the intermediate temperature was larger than that at the low temperature when the magnetic field was intermediate. This phenomenon disappears when the magnetic field is higher than the anisotropy field at a low temperature. In the case of a magnetic field equal to 50 kOe, a difference can be observed in the slope of magnetization between the easy and hard planes. It only appears at a specific temperature range from the intermediate temperature to slightly above the Curie temperature. This is semi-quantitative in agreement with the experimental results\cite{McGuireCM2015,LiuPRB2018}.

%para 10
From the $M$--$H$ curves, the temperature dependence of the MAE was estimated as FIG. \ref{FIG3} (c) by considering the area between the easy axis and the hard plane. The value of MAE at low temperatures is approximately equal to the value of first-principles calculations (0.46 meV/Cr-atom). The MAE decreased almost linearly when the temperature increased in the low-temperature range. The MAE curve becomes flatter at higher temperatures, leading to a finite value slightly above the Curie temperature. This is due to magnetization anisotropy and anisotropic magnetic susceptibility, as discussed above. This means that the magnetic anisotropy is still valid at a temperature higher than the critical temperature, at which the anisotropic magnetic susceptibility and magnetization anisotropy are present. The effect of MAE was proposed as the origin of the anisotropic magnetocaloric effect in experimental studies\cite{LiuPRB2018}. This means that the effect of MAE is not negligible slightly above the Curie temperature in the present study.

%para 11
The isothermal magnetization curves and isothermal magnetic entropy changes are shown in FIG. \ref{FIG4}. The difference in magnetization and entropy change is due to the effect of the MAE. The magnetization at the zero external magnetic field is approximately zero for the hard plane case because the easy axis is perpendicular to the hard plane in uniaxial anisotropy. The magnetization in the low magnetic-field and low temperature region is affected by the magnetic field and MAE in the case of hard plane. The higher applied magnetic field, the higher magnetization value. When the magnetic field is higher than the anisotropy field, the magnetization is perpendicular to the easy axis. This leads to the disappearance of MAE effects for a sufficient magnetic field applied. The magnetization curves become smooth after this region. If the strength of the external magnetic field increases, the magnetization can be rotated to the hard axis. The entropy change of the hard plane is quite different from that in the easy axis case. Here, the entropy of the external magnetic field along the easy axis is shown as a positive peak, where all values have a positive sign. Meanwhile, the entropy change of the hard plane crosses the zero line. The entropy change is negative at a magnetic field of 1 T before the Curie temperature peak. When the magnetic field increases, the crossing temperature decreases. This means that the material will be cooled by applying a magnetic field along the hard plane in the adiabatic process for this temperature range. This behavior is unusual because conventional magnetocaloric materials are heated and cooled by applying and removing the external magnetic field in the adiabatic process\cite{McGuireCM2015,LiuPRB2018}. The origin of the negative sign in the entropy change curve of the hard plane comes from the peak, as in the magnetization curve at intermediate temperatures and insufficient magnetic fields. MAE tends to force the magnetization along the easy axis, whereas the magnetic field is along the hard plane. Although the external magnetic field is constant, the strength of MAE decreases when the temperature increases, which leads to a reduction in the anisotropy field. If the MAE effect are weakened by temperature, the magnetization can be rotated easily, leading to higher magnetization along the direction of the magnetic field. Our results are in good agreement with the experimental works\cite{McGuireCM2015,LiuPRB2018}.

%para 12
The entropy change as a function of the magnetic field applied to the easy axis, and the hard plane at some temperatures is shown in FIG. \ref{FIG5} (a) and (b). Note that the Curie temperature of our present work is higher than the experimental value, so we use higher temperature to keep the normalize temperature similar to that of the experimental works\cite{LiuPRB2018}. The entropy change curve of the easy axis and hard plane at the same temperature is distinguished in the entire region of the applied magnetic field (up to 7.4 T). The entropy change of the easy axis at all temperatures is positive and increases almost linearly with the magnetic field value. At the considered temperature, a higher temperature leads to a steeper slope of the entropy change. In contrast, when the temperature is smaller than the Curie temperature, the entropy change of the hard plane (the solid line) exhibits similar behavior. The entropy change decreases when the magnetic field increases and reaches a minimum and then increases with a similar slope of the easy axis. The position of the minimum entropy change decreases when the temperature increase. The distance between the two entropy change curves (dashed and solid lines) increases when the magnetic field is smaller than the minimum and becomes constant at a sufficient applied magnetic field. Here, a temperature slightly higher than the Curie temperature is also considered with yellow line in FIG. \ref{FIG5} (a) and (b). At this temperature, there is no minimum in the entropy change curve of the hard plane. In addition, the slopes of the two curves is different over the entire magnetic field range. This means that there is a difference in the entropy change between the hard plane and easy axis at slightly above Curie temperature.  

%para 13
The magnetic susceptibility of the hard plane case was considered in FIG. \ref{FIG5} (c) to understand the origin of the behavior of the entropy change. The boundary of the two regions (the green and blue colors), which correspond to insufficient and sufficient magnetic fields, can be clearly seen by considering the magnetic susceptibility. Magnetic susceptibility is the fluctuation of the magnetization value. It is large in the low temperature and low magnetic field region, where there is competition between the magnetic field and magnetic anisotropy. If the magnetic field is sufficiently higher than the anisotropy field, the effect of MAE on the magnetization and magnetic susceptibility becomes inferior because the direction of magnetization is perpendicular to the easy axis. The boundary of the two regions in magnetic susceptibility corresponds to the strength of the magnetic field of the minimum entropy change in the hard plane case in FIG. \ref{FIG5} (a) and (b). It is high at low temperatures and decreases when the temperature increases and disappears above the Curie temperature.

\subsection{Relations of MAE and anisotropic magnetocaloric effect}

%para 14
To understand the relation between MAE and entropy change of the magnetocaloric effect, we define the anisotropic magnetic entropy as the heat absorbed or released by rotating the sample at a fixed strength of the magnetic field in the isothermal process as:

\begin{equation}
\begin{split}
&\Delta S^{\rm R}_{\rm M}(H_{\rm ext}, T)\\ 
&=S(H_{\rm ext}\parallel{\rm easy},T)-S(H_{\rm ext}\parallel{\rm hard},T)\\
%&=S(H_{\rm ext}\parallel{\rm axis\;1},T)-S(0,T)+S(0,T)-S(H_{\rm ext}\parallel{\rm axis\;2},T)\\
&=\Delta S_{\rm M}(H_{\rm ext}\parallel{\rm easy},T)-\Delta S_{\rm M}(H_{\rm ext}\parallel{\rm hard},T)\\
&=\int_{0}^{H_{\rm ext}} \left \{ \frac{\partial M(H\parallel{\rm easy},T) }{\partial T} - \frac{\partial M(H\parallel{\rm hard},T) }{\partial T} \right \}, \\
&=\frac {\partial E_{\rm MAE}}{\partial T}
\end{split}
\label{Eq6}
\end{equation}

%para 15
The anisotropic magnetic entropy is related to the derivative of MAE, as shown in Eq. \ref{Eq6}. This means that the difference in entropy at a given temperature is slightly higher than that at the Curie temperature as yellow curves in FIG. \ref{FIG5} (a, b) is consistent with the finite MAE at these temperatures of FIG. \ref{FIG3} (c). In addition, at temperatures below the Curie temperature, the estimation of MAE increases as the magnetic field increases and becomes approximately constant if the magnetic field is larger than the anisotropy field. This means that the deepness of the minimum entropy change of the hard plane is directly related to the strength of the MAE, which decreases when the temperature increases. The MAE reduction at a finite temperature is related to the anisotropic entropy because the MAE is the difference in free energy, which includes the internal energy and entropy terms. Although our methods can successfully reproduce the experimental results and understand their characterization, a further study using these methods on cubic anisotropy and two ion anisotropy cases, is needed to overview MAE in the magnetocaloric effects.

\section{\label{sec:level4}Summary}
The electronic structure, magnetism, and magnetocaloric properties of CrI$_{3}$ were studied by combining first-principles calculations and Monte--Carlo simulations. The temperature dependence of the MAE, estimated by integrating the $M$--$H$ curves as the Sucksmith-Thompson method, is not negligible even above the Curie temperature. Its origin is anisotropic magnetic susceptibility and magnetization anisotropy, which are considered in our Monte--Carlo simulations. The relationship between anisotropic magnetic entropy change and finite temperature magnetic anisotropy energy is clarified in this study to understand the mechanism of the anisotropic magnetocaloric effect. Base on our results, the anisotropic magnetic entropy is the main source of temperature dependence of magnetic anisotropy energy where the entropy terms appears in the formulation of free energy.

%\section*{CRediT authorship contribution statement}
%H. B. Tran: Investigation, Writing - Original draft preparation. H. Momida: Writing - Reviewing and Editing, Supervision. K. Sato: Writing - Reviewing and Editing. Y. Matsushita: Writing - Reviewing and Editing. Y. Makino: Writing - Reviewing and Editing. T. Oguchi: Writing - Reviewing and Editing, Supervision. 

\section*{Acknowledgements}
The authors would like to thank Tetsuya Fukushima for his help with first-principles calculations. The authors would also like to thank Kunihiko Yamauchi for the valuable discussions. This work was partly supported by the Di-CHiLD project (Grant No. J205101520). The computation in this work was partially performed using the facilities of the Supercomputer Center, the Institute for Solid State Physics, the University of Tokyo, and the Cybermedia Center of Osaka University.

%\section*{Data availability}
%The raw data required to reproduce these findings are available by making an e-mail request to the corresponding author. 

%========References============================
\bibliography{basename of .bib file}

\end{document}